\documentclass[twocolumn,showpacs,preprintnumbers,amsmath,amssymb,superscriptaddress]{revtex4-1}
\usepackage{graphicx}
\graphicspath{ {figs/} }

\newcommand{\gsim}{\;\rlap{\lower 3.5 pt \hbox{$\mathchar \sim$}} \raise 1pt
 \hbox {$>$}\;}
\newcommand{\lsim}{\;\rlap{\lower 3.5 pt \hbox{$\mathchar \sim$}} \raise 1pt
 \hbox {$<$}\;}

\usepackage{color}

\begin{document}

\title{\boldmath
Next-to-next-to-next-to-leading order QCD prediction for
the top anti-top S-wave pair production cross section near threshold 
in $e^+ e^-$ annihilation}

\author{Martin Beneke}
\affiliation{Physik Department T31, James-Franck-Stra\ss{}e~1,
    Technische Universit\"at M\"unchen, D-85748 Garching, Germany}

\author{Yuichiro Kiyo}
\affiliation{Department of Physics, Juntendo University,
Inzai, Chiba 270-1695, Japan}

\author{Peter Marquard}
\affiliation{Deutsches Elektronen Synchrotron DESY,
      Platanenallee 6, D-15738 Zeuthen, Germany}

\author{Alexander Penin}
\affiliation{Department of Physics, 
 University of Alberta,
  Edmonton AB T6G 2J1, Canada}
\affiliation{Institut f{\"u}r Theoretische Teilchenphysik, Karlsruhe
  Institute of Technology (KIT), 76128 Karlsruhe, Germany}

\author{Jan Piclum}
\affiliation{
Albert Einstein Center for Fundamental Physics,
Institute for Theoretical Physics,
Sidlerstrasse 5,
CH-3012 Bern, Switzerland}

\author{Matthias Steinhauser}
\affiliation{Institut f{\"u}r Theoretische Teilchenphysik, Karlsruhe
  Institute of Technology (KIT), 76128 Karlsruhe, Germany}

\date{June 17, 2015}

\begin{abstract}
\noindent 
  We present the third-order QCD prediction for the production of 
  top-anti-top quark pairs in electron-positron collisions close to 
  the threshold in the dominant S-wave state. We observe a significant 
  reduction of the theoretical uncertainty and discuss the sensitivity 
  to the top quark mass and width.
\end{abstract}
 
\preprint{
  ALBERTA-THY-12-15,
  DESY-15-100,
  SFB/CPP-14-127,
  TTP15-021,
  TUM-HEP-1001/15}
 
\maketitle

\noindent
Among the main motivations for building a future high-energy 
electron-positron collider 
in steps of increasing center-of-mass energy are precise measurements at 
$\sqrt{s}\approx 345\,$GeV close to the production threshold of top anti-top 
quark pairs.  The peculiar behaviour of the cross section
allows for the precise determination of a number of Standard Model parameters,
most prominently the top quark mass.  To date the most precise measurement of
$m_t = 173.34\pm 0.27\mbox{(stat)}\pm 0.71\mbox{(syst)}\,$GeV 
comes from the hadron colliders 
Fermilab Tevatron and CERN LHC~\cite{ATLAS:2014wva} and  
is based on the reconstruction of the top and
anti-top quarks through their decay products. 
This approach and the value 
quoted above are plagued by unknown relations of the extracted mass value 
$m_t$ to top quark masses in the pole
or $\overline{\rm MS}$ renormalization scheme, which may well 
exceed 1~GeV. At hadron colliders there are
also methods to determine directly 
a well-defined top quark mass, such as the extraction of
$m_t$ from top quark cross section measurements.  However, the final precision 
is of the order of a few GeV and thus significantly worse.  At an 
electron-positron collider, on the other hand, scans of the top anti-top 
pair production threshold can lead to very precise measurements of 
well-defined mass values with a statistical accuracy of  
only 20-30~MeV~\cite{Seidel:2013sqa,Horiguchi:2013wra}.
Besides the top quark mass also its decay width and the
strong coupling constant can be extracted with an accuracy of
21~MeV~\cite{Horiguchi:2013wra} and 0.0009~\cite{Seidel:2013sqa},
respectively. A recent study has shown that for 
a Higgs
boson with a mass of about 125~GeV the top quark Yukawa coupling can be
obtained with a statistical uncertainty of only
4.2\%~\cite{Horiguchi:2013wra}.  
These numbers pose several challenges to theory.

A crucial input to reach the aimed precision is a precise calculation of the
top anti-top pair production cross section in the threshold region.  
While the fundamental theory of quantum chromodyanmics (QCD) 
is well-established,
performing calculations of quantum corrections to the very high 
accuracies demanded here is very difficult indeed. The problem is 
further complicated by the fact that in the threshold region 
the colour Coulomb potential 
$\propto \alpha_s/r$, where $\alpha_s$ denotes the 
strong coupling, can no longer be treated as a perturbation even 
though $\alpha_s\ll 1$. Standard perturbation theory in $\alpha_s$ 
breaks down and resummation is required.

The relevant techniques have been developed in the 1990s in the 
framework of effective field theory (EFT), which accounts for the
different dynamical scales in the problem.  For a heavy quark anti-quark 
system at threshold there are three relevant scales, the hard scale $m$, 
the potential and soft scale $m v$, and the ultrasoft scale $mv^2$, 
where $m$ denotes the mass of the quark and $v$ its velocity. 
Since $v \ll 1$ there is indeed a strong hierarchy between these scales
and thus it is possible to construct a tower of effective theories taking QCD
as starting point.  In a first step one arrives at non-relativistic QCD
(NRQCD)~\cite{Thacker:1990bm,Lepage:1992tx}
by integrating out the hard modes. Afterwards all other modes are integrated 
out except the ones which are needed to describe a physical 
non-relativistic quark anti-quark system---potential modes for the 
quarks with energy ${\cal O}(mv^2)$ and three-momentum ${\cal O}(mv)$, 
and ultrasoft gluons with four-momentum ${\cal O}(mv^2)$.  
The corresponding theory is potential NRQCD (PNRQCD)~\cite{Pineda:1997bj}. 
In practice, the separation of the different modes and scales is 
done with the threshold expansion of Feynman diagrams~\cite{Beneke:1997zp}. 
The concepts and tools required to perform the computation of the 
cross section near threshold with the accuracy reported in this 
Letter are summarized in~\cite{Beneke:2013jia}.  

Within the EFT the normalized total cross section can be written in the
form (see, e.g.,~\cite{Beneke:2013jia}) 
\begin{eqnarray}
  R &=& \frac{\sigma(e^+ e^- \to t\bar{t} +X)}{\sigma_0}
\\
  &=& \frac{18\pi}{m_t^2} \,\mbox{Im}\left\{ 
    c_v \left[
      c_v-\frac{E}{m_t}\left(c_v+\frac{d_v}{3}\right)
    \right] G(E) + \ldots
  \right\}
  \,,
  \nonumber
\end{eqnarray}
where $c_v$ and $d_v$ denote NRQCD matching coefficients, 
$E=\sqrt{s}-2m_t$, $G(E)$ is the non-relativistic two-point Green function, 
and $\sigma_0=4\pi\alpha^2/(3s)$ is the cross section for the production of 
a $\mu^+\mu^-$ pair in the limit of large center-of-mass energy $\sqrt{s}$. 
The ellipsis refers to terms which are beyond the 
next-to-next-to-next-to-leading order (N$^3$LO) in the expansion in 
$\alpha_s$ and $v$.

The next-to-next-to-leading order (NNLO) QCD corrections were computed 
in the late 1990s~\cite{Hoang:1998xf,Melnikov:1998pr,Penin:1998ik,Beneke:1999qg,Hoang:1999zc,Nagano:1999nw}, and the results of several groups 
were summarized and compared in~\cite{Hoang:2000yr}. The first NNLO 
calculations were expressed in terms of the top quark pole mass 
and found large corrections to the location of the cross section peak 
near threshold casting doubt on the possibility to perform a very accurate 
mass measurement. It was pointed out in~\cite{Beneke:1998jj} that 
these corrections are an artifact of the renormalization convention, 
which could be avoided by choosing a scheme that is less sensitive 
to uncalculable long-distance effects of QCD. Subsequent calculations 
of the top anti-top threshold all employ mass renormalization conventions 
different from the on-shell scheme. Here we use the 
potential-subtracted (PS) mass \cite{Beneke:1998rk}. However, even 
with this improvement, large corrections to the height of the cross 
section peak are observed at NNLO, which motivates the N$^3$LO 
QCD calculation, the result of which is reported in this Letter.

There has been quite some effort to resum logarithmic terms in the 
velocity of the produced top quarks and obtain
so-called next-to-next-to-leading logarithmic
(NNLL) approximations~\cite{Hoang:2001mm,Pineda:2006ri}. The most complete
analysis has been performed in~\cite{Hoang:2013uda}, where new ultrasoft
terms have been included. The NNLL approximation already contains the 
$\ln v$ enhanced terms of the N$^3$LO correction, but not the 
``constant'' terms. Partial results have shown that these constant 
terms are as large as the logarithmic terms, at least in individual 
pieces of the 
calculation \cite{Beneke:2008cr,Marquard:2014pea,BenKiy_II}.

Since more than ten years several groups have computed building
blocks of the N$^3$LO correction, which can be subdivided into 
matching coefficients of NRQCD and PNRQCD
and higher order corrections to $G(E)$.  In~\cite{Beneke:2013jia} a detailed
discussion of the individual contributions can be found. Among them are the
three-loop corrections to the static
potential~\cite{Smirnov:2008pn,Smirnov:2009fh,Anzai:2009tm}, certain 
${\cal O}(\epsilon)$ terms of higher-order potentials in dimensional 
regularization~\cite{Beneke:2013jia,Wuester:2003,Beneke:2014qea},
and three-loop fermionic
corrections to $c_v$~\cite{Marquard:2009bj,Marquard:2006qi}. Furthermore,
ultrasoft corrections to $G(E)$~\cite{Beneke:2008cr} and all 
Coulombic contributions up to the third order~\cite{Beneke:2005hg}. 
Recently, the last two building blocks have been computed, which 
allows us to put together for the first time a prediction 
for the cross section near threshold with N$^3$LO accuracy. First, 
the purely gluonic three-loop correction to the QCD-to-NRQCD 
vector current matching 
coefficient $c_v$ appearing in 
\begin{eqnarray}
\label{eq:QCDVectorCurrent}
\bar t\gamma^i t =c_v\, \psi^\dag\sigma^i\chi
+ \frac{d_v}{6m_t^2}\,\psi^\dag\sigma^i\,{\bf D^2}\chi
+\ldots,
\end{eqnarray}
has been computed \cite{Marquard:2014pea}, completing the matching 
calculations up to the so-called singlet diagrams. (In other contexts  
and at second order \cite{Kniehl:2006qw} these singlet contributions 
have been shown to be small.) 
Second, the calculation of the PNRQCD two-point function 
$G(E)$ of currents $\psi^\dag\sigma^i\chi$ has been completed to third order 
in PNRQCD perturbation theory with the computation 
of the single and double non-Coulomb potential insertions~\cite{BenKiy_II}. 
We note that PNRQCD includes the leading-order Coulomb potential 
in the unperturbed Lagrangian. Perturbation theory around this 
interacting Lagrangian amounts to the resummation of the standard 
loop expansion in $\alpha_s$. 
The various contributions have been encoded in a {\tt Mathematica} 
program. 

There is a number of non-QCD effects, which are expected to be 
relevant to a realistic cross section prediction. a)~The existence
of a Higgs boson affects the production process and generates 
an additional short-range potential. b)~The same holds for electroweak 
and electromagnetic corrections, amongst which is the electrostatic 
potential between the charged top quarks. c)~The rapid electroweak 
decay $t\to W^+ b$ of the top quark is responsible for the fact 
that the distinct toponium bound-state poles are smeared to a broad 
resonance near threshold (see  Fig.~\ref{fig::mu} below). For the 
same reason, it will be mandatory to consider the process 
$e^+ e^-\to W^+ W^-b\bar b$ including the non-resonant contributions. 
Non-resonant effects have been computed to NLO and partially to 
NNLO accuracy \cite{Beneke:2010mp,Penin:2011gg,Jantzen:2013gpa,Ruiz-Femenia:2014ava}, and can be naturally separated in the framework of unstable 
particle EFT \cite{Beneke:2003xh,Beneke:2004km}. d)~Electromagnetic 
initial-state radiation generates large logarithms $\ln(4 m_t^2/m_e^2)$, 
where $m_e$ is the electron mass, which must be summed. e)~Furthermore, 
there are collider-specific effects such as beamstrahlung and the 
luminosity spectrum~\cite{Seidel:2013sqa}. Neither of these effects 
is discussed further in the following. The purpose of the present 
Letter is to demonstrate that the third-order QCD calculation brings 
the QCD contributions under control, which is a prerequisite for all 
further studies, including the non-QCD effects above. We refer to 
\cite{BMPR} for a study of the Higgs contribution and the prospects for 
determining the Yukawa coupling, and the impact of the NLO non-resonant 
correction on the top mass determination.

We now turn to the discussion of the N$^3$LO QCD result. In order 
to better compare the successive orders we do not include here the 
known contribution \cite{Beneke:2013kia} from the axial-vector 
$Z$-boson coupling in the production process $e^+ e^- \to 
Z^*\to t\bar {t}$, which starts at NNLO and is below 1\%. Our 
result therefore refers to the S-wave production cross section.
Unless stated otherwise we use the following input values
for the top quark mass in the PS scheme, the top quark width
and the strong coupling constant:
\begin{eqnarray}
  && m_t^{\rm PS}(\mu_f=20\,\mbox{GeV}) = 171.5~\mbox{GeV},\,
  \Gamma_t = 1.33~\mbox{GeV},
  \nonumber\\
  && \alpha_s(M_Z) = 0.1185 \pm 0.0006
  \,.
  \label{eq::input}
\end{eqnarray}
Furthermore, the renormalization scale is varied between 50 and 350~GeV.
Note that an unstable behaviour of the perturbative series is expected 
for scales below $\mu\approx40$~GeV~\cite{BenKiy_II,Beneke:2005hg}. The scale 
$\mu_w$ related to the separation of resonant and non-resonant 
contributions to the $e^+ e^-\to W^+ W^-b\bar b$ process is fixed 
to $\mu_w = 350\,$GeV. The dependence on this choice is numerically 
negligible.

\begin{figure}[th]
  \centering
  \hskip-0.8cm
  \includegraphics[width=.9\linewidth]{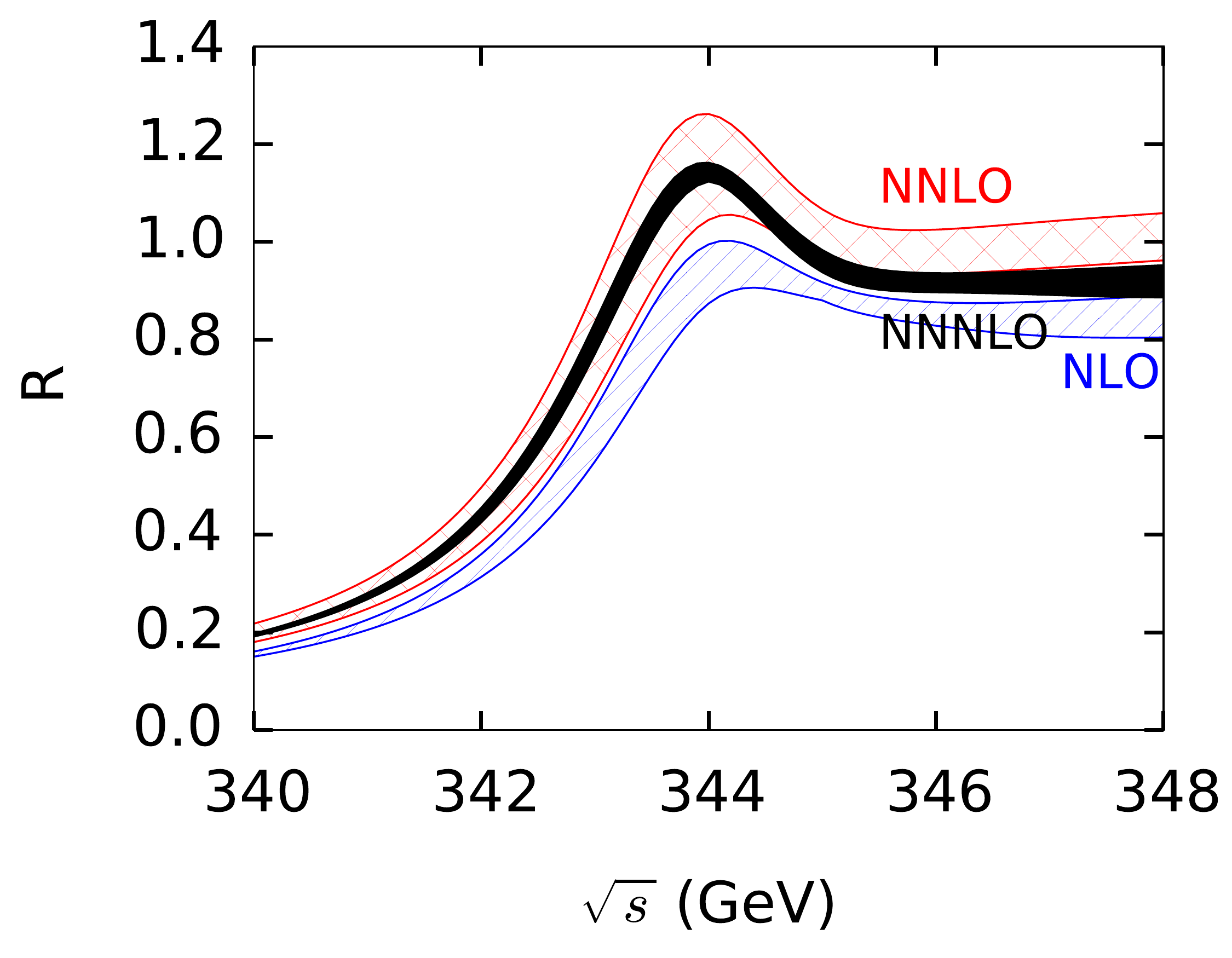}
  \caption{Scale dependence of the cross section near threshold. The NLO,
    NNLO and N$^3$LO result is shown in blue, red and black,
    respectively. The renormalization scale is varied between $50$ and
    $350$~GeV.}
  \label{fig::mu}
\end{figure}

The main result of this Letter is shown in Fig.~\ref{fig::mu},
where the total cross section is shown as a function of the center-of-mass
energy $\sqrt{s}$. The previous NLO 
and NNLO predictions are also shown for comparison to the new N$^3$LO 
result (black, solid). The bands are obtained by variation of 
the renormalization scale in the specified range.
After the inclusion of the third-order corrections one observes a dramatic
stabilization of the perturbative prediction, in particular in and below the
peak region. In fact, the N$^3$LO curve is entirely contained within the 
NNLO one. This is different above the peak position where a clear negative 
correction is observed when going from NNLO to N$^3$LO. For example, 3~GeV 
above the peak this amounts to $-8\%$. This arises from the large negative 
three-loop correction to the matching coefficient 
$c_v$~\cite{Marquard:2014pea}. 

\begin{figure}[b]
  \centering
  \hskip-0.5cm
  \includegraphics[width=.9\linewidth]{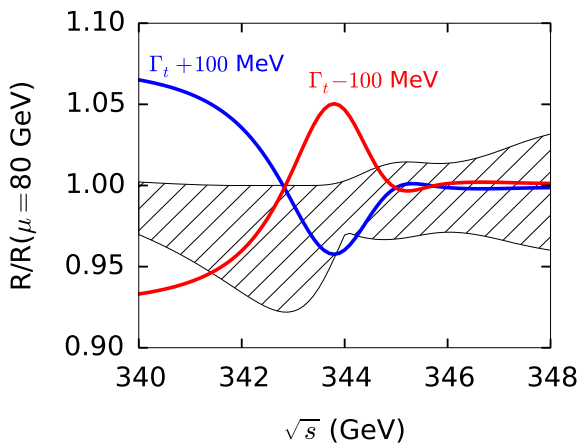}  
  \caption{Scale dependence  (hatched area) 
of the N$^3$LO cross section relative to the 
reference prediction. Overlaid are predictions for two 
different values of $\Gamma_t$, again normalized to the 
reference prediction. See text for details.}
\label{fig::gamma}
\end{figure}

The theoretical precision of the third-order QCD result as measured by the 
residual scale dependence is highlighted in Fig.~\ref{fig::gamma}, which 
shows $R(\mu)$ normalized to a reference prediction defined at $\mu=80\,$GeV. 
The width of the shaded band corresponds to an uncertainty of 
about $\pm 3\%$ with some dependence on the center-of-mass energy 
$\sqrt{s}$. The figure also shows the sensitivity to the top-quark 
width. The two solid lines refer to the cross section with 
$\Gamma_t$ changed by $\pm 100\,$MeV to $1.43$ and $1.23\,$GeV, 
respectively, computed with $\mu=80\,$GeV and normalized to the reference 
prediction. Decreasing the width implies a sharper peak, i.e.~an 
enhancement in the peak region, and a suppression towards the 
non-resonant region below the peak. A few GeV above the peak the 
cross section is largely insensitive to the width. Increasing the width 
leads to the opposite effects. This pattern is clearly seen in 
Fig.~\ref{fig::gamma}, which also demonstrates that a $\pm 100\,$MeV 
deviation from the width predicted in the Standard Model leads to a 
cross section change near and below the peak far larger than the 
uncertainty from scale variation.

\begin{figure}[b]
\centering
\hskip-0.4cm
\includegraphics[width=0.9\linewidth]{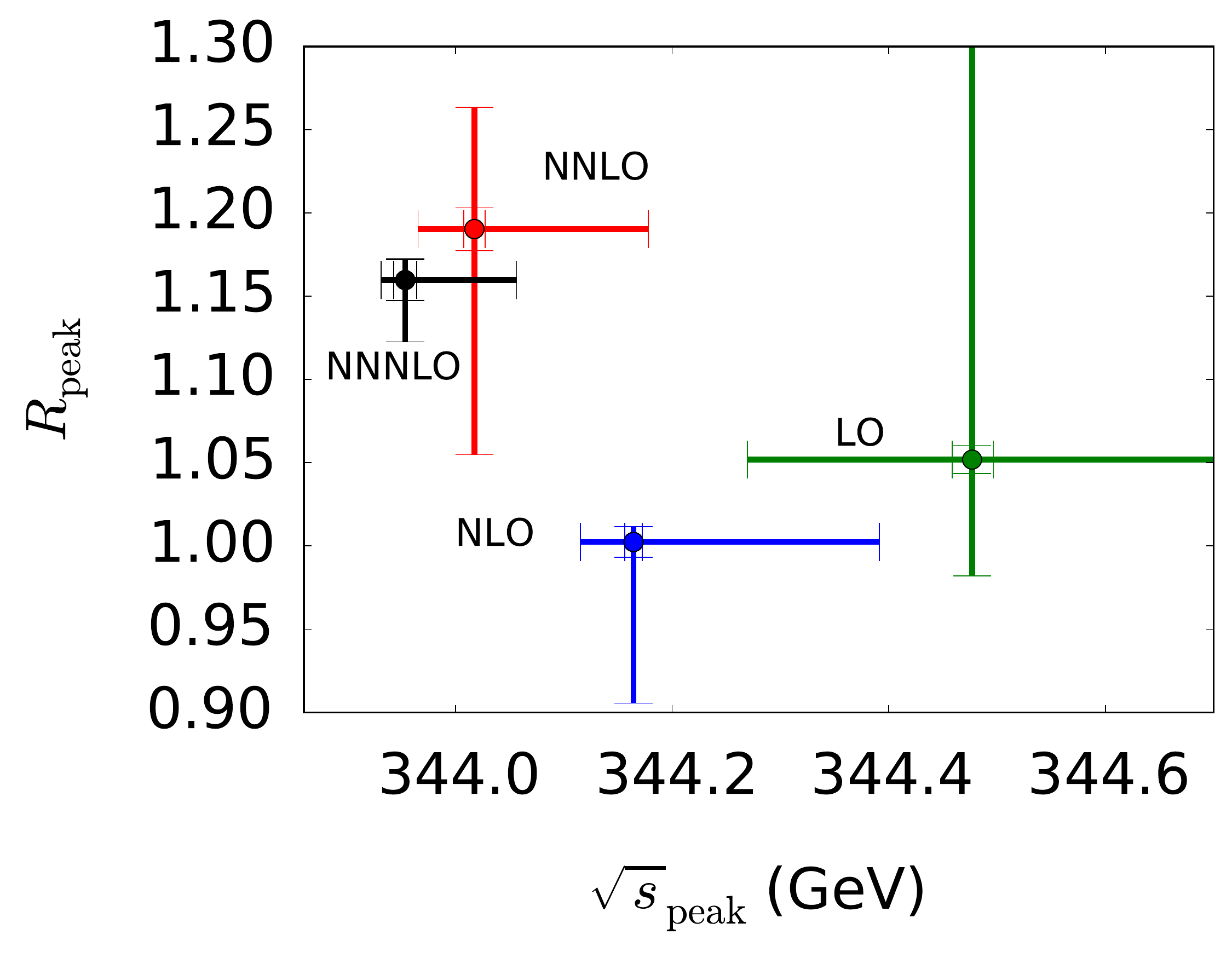}    
\caption{\label{fig::peak}
Position and height of the cross section peak at LO, NLO, NNLO and 
N$^3$LO. The unbounded range of the LO error bars to the right and up 
are due to the fact that the peak disappears for large values of the 
renormalization scale.}
\end{figure}

We now turn to the question to what accuracy the top quark mass can 
be determined. Even if we focus only on the theoretical accuracy, a 
rigorous analysis requires accounting for the specifics of the energy 
points of the threshold scan and the correlations. However, a good 
indication is already provided by looking at the position and height of the
resonance peak.  Fig.~\ref{fig::peak} shows this information at LO, NLO, 
NNLO and N$^3$LO, where the outer error bar reflects the uncertainty 
due to the renormalization scale and $\alpha_s$ variation, 
added in quadrature, and the inner error bar only takes the $\alpha_s$  
uncertainty into account.  The central point refers to the value at the 
reference scale $\mu=80\,$GeV. There is a
relatively big jump from LO to NLO of about $310$~MeV, approximately 150~MeV 
from NLO to NNLO, which reduces to only 64~MeV from NNLO to N$^3$LO. 
Furthermore, the NNLO and N$^3$LO uncertainty bars show a significant 
overlap. Taking into account only the uncertainty from scale variation 
the uncertainty of the peak position amounts to $\pm 60$~MeV at N$^3$LO with 
a factor two improvement relative to NNLO. 
The improvement is even larger for the peak height, which is relevant to 
the top quark width and Yukawa coupling determination as discussed above 
and in \cite{BMPR}. Note that these conclusions refer to the top quark 
PS mass (and not the pole mass), and correspondingly to the 
$\overline{\rm MS}$ mass, which can be related to the PS mass with an 
accuracy of about 20~MeV~\cite{Marquard:2015qpa}.

\begin{figure}[t]
  \centering
  \hskip-0.3cm
  \includegraphics[width=0.95\linewidth]{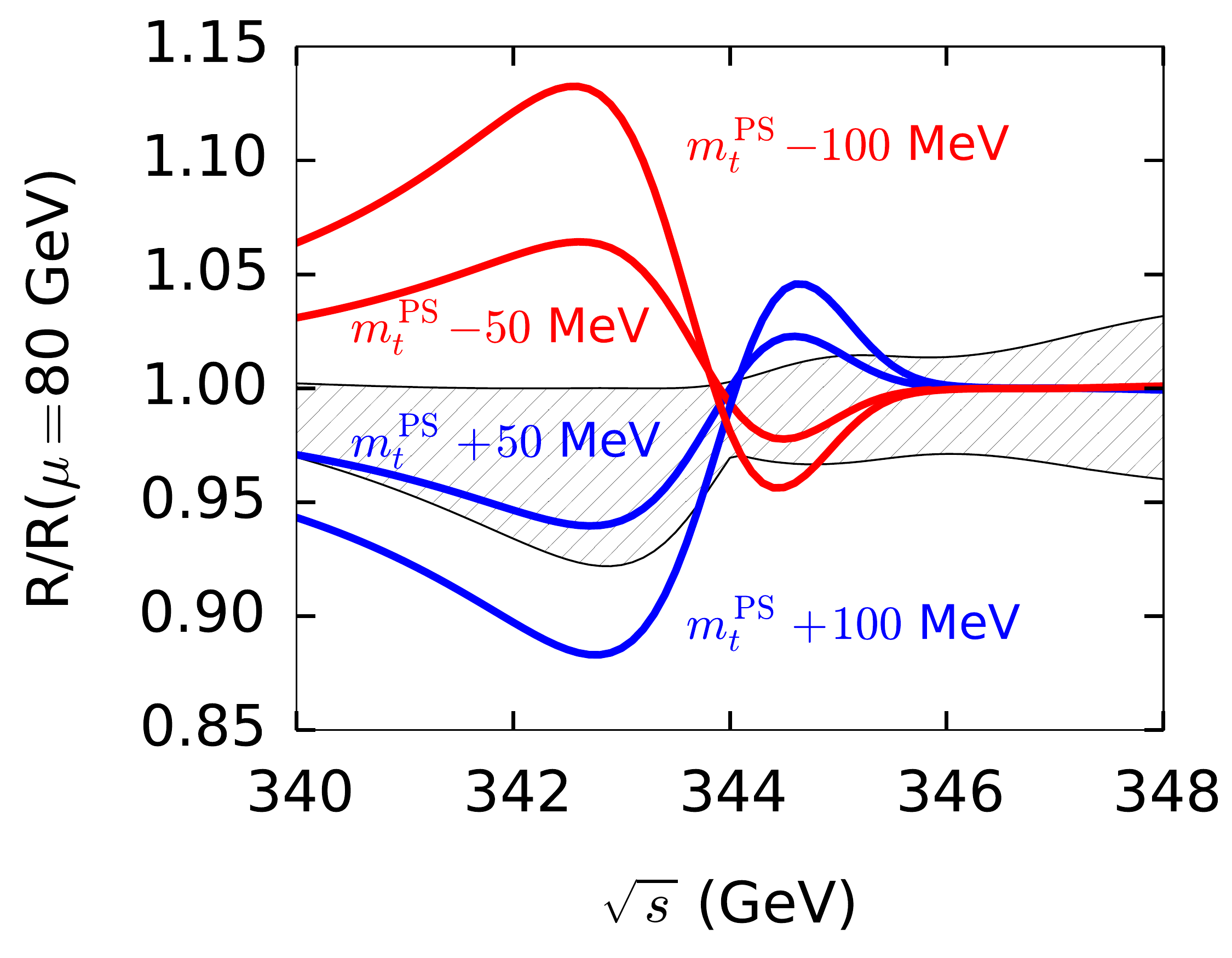}    
\caption{\label{fig::norm}
Relative cross section variation when the top quark mass is 
changed by $\pm 50\,$ and $\pm 100\,$MeV, superimposed on the 
scale dependence of the N$^3$LO cross section  
prediction. See text for details.}
\end{figure}

We display the sensitivity of the cross section to the top quark mass 
in Fig.~\ref{fig::norm}, which is the same as Fig.~\ref{fig::gamma}, except 
that the curves superimposed to the relative scale variation of the 
N$^3$LO cross section now refer to shifts of the top PS mass by 
$\pm 50\,$ and $\pm 100\,$MeV. The shape of these curves is easily 
understood, as a change of the mass by $\delta m_t$ mainly shifts the 
peak of the cross section and is largely equivalent to a shift of 
$\sqrt{s}$ by $2\delta m_t$. The figure demonstrates that the largest 
sensitivity to the mass occurs around $1.5\,$GeV below the peak. 
A variation of the mass by $\pm 50\,$MeV changes the cross section 
at $\sqrt{s}=342.5\,$GeV by 
$\pm 6\%$, compared to a scale uncertainty of $\pm 3.8\%$. 
Accounting for the characteristic shape of variation, we may conclude 
that the theoretical uncertainty on the top quark mass should be 
well below $\pm 50\,$MeV. 

Comparing the N$^3$LO result to the NNLO 
results with $\ln v$ resummation~\cite{Pineda:2006ri,Hoang:2013uda}, 
we notice that in both approaches the correction to the peak height 
relative to NNLO  is negative (with the default scale 
choice)~\cite{Pineda:2006ri}, 
while the theoretical uncertainty of the cross section normalization is 
reduced to about $\pm 3\%$ compared to the $\pm 5\%$ quoted 
in~\cite{Hoang:2013uda}.

In summary, we presented the third-order QCD calculation of the 
top quark production cross section in $e^+ e^-$ annihilation in the 
threshold region. Despite the extra complication of non-relativistic 
resummation to all orders in QCD perturbation theory, this is one of 
only a few collider processes now known at N$^3$LO in QCD. The third-order 
calculation leads to a large reduction of the QCD theoretical 
uncertainty to about $\pm 3\%$, and thereby solves a long-standing 
issue regarding the reliability of the QCD prediction for this process. 
Top quark mass determinations with theoretical errors below 
$50\,$MeV now appear feasible. Further improvement by going to the 
next order in QCD is currently unrealistic. However, it would be desirable 
to include the information about logarithmic effects in $v$ beyond 
N$^3$LO already contained in the NNLL 
computations~\cite{Pineda:2006ri,Hoang:2013uda}. More importantly, 
with QCD effects under control as demonstrated here, further studies 
of non-QCD effects are now well motivated---and required.

\section*{Acknowledgments}

\noindent 
This work has been supported by DFG Sonderforschungsbereich Transregio 9 
``Computergest\"utzte Theoretische Teilchenphysik'', 
and the DFG Gottfried Wilhelm Leibniz programme. The work of Y.K. 
was supported in part by Grant-in-Aid for Scientific 
Research No. 26400255 from  MEXT, Japan. P.M. was supported in part 
by the European Commission through contract PITN-GA-2012-316704 (HIGGSTOOLS).
The work of A.P. is partially supported by NSERC.

\end{document}